\documentclass[%
 reprint,
 amsmath,amssymb,
 aps,
]{revtex4-2}

\usepackage{graphicx}
\usepackage{dcolumn}
\usepackage{bm}
\usepackage{color}

\begin{document}

\preprint{APS/123-QED}

\title{Burstiness and information spreading in the active particles systems}

\author{Wei Zhong$^1$} 
\email{w.zhong@mju.edu.cn}
\author{Youjin Deng$^{1,2}$}
\author{Daxing Xiong$^1$}
\email{xmuxdx@163.com}
\affiliation{$^1$ MinJiang Collaborative Center for Theoretical Physics, College of Physics and Electronic Information Engineering, Minjiang University, Fuzhou 350108, P. R. China.\\
$^2$ Department of Modern Physics, University of Science and Technology of China, Hefei, Anhui 230026, China.
}%




\date{\today}

\begin{abstract}
We construct the temporal network using the two-dimensional active particle systems which are described by the Vicsek model. The bursts of the interevent times for a specific pair of particles are investigated numerically. We find that for different noise strength, the distribution of the interevent times of a target edge follows by a heavy-tail, revealing a strong burstiness of the signals. To further characterize the nature of the burstiness, the burstiness parameter and the memory coefficient are calculated. The results show that near the critical points of the Vicsek model, the burstiness parameters reach the minimum values for each density, indicating a relation between the phase transition of the Vicsek model and the bursty nature of the signals. Besides, the memory plays a negligible role in the burstiness. Further, we investigate the spreading dynamics on our temporal network with the susceptible-infected model, and observe a positive correlation between the burstiness and the information spreading dynamics. 
\end{abstract}

\maketitle

\section{Introduction \label{sec1}}
The temporal networks, with the links vary with time, attract more and more attentions in the last decades \cite{holme2012,holme2015,li2017} . 
Unlike in the static networks, where the structural heterogeneities is important, the burstiness, corresponding to a power-law distribution of the interevent time, plays an important role in the temporal networks. The burstiness has been observed in a variaty of networks \cite{bara,gold,moin}. 

Goh and Barab\'asi \cite{goh2008} reported that the bursty nature of a signal can be normally described by two distinct quantities: 
one is the interevent time distribution, which can be characterized by the burstiness parameter $B$; and the other quantity is the memory, usually described by the memory coefficient $M$. With the ($M,B$) phase diagram, Goh and Brab\'asi successfully distinguished the origin of the bursty nature for a wide variety of signals from human activities, to natural phenomena, texts, and heart beat, {\it etc}. However, for a time series, normally, its length is normally finite and can strongly affect the measurement of the burstiness parameter. Therefore, Kim and Jo \cite{kim2016} proposed a finite-size formula of the burstiness parameter, which reduced the finite-size effects.  

When the bursty nature of a signal is identified, it can produce aging effects, resulting in the breaking of the time translation invariance of the degree distribution, in the corresponding time-integrated network \cite{moinet2015}. Besides, it can also suppress or promote the spreading dynamics on the temporal networks \cite{karsai2011,lambiotte2013,horvath2014,
perotti2014,tizzani2018,michalski2020}. 
Vazquz {\it et al.} \cite{vazquz2007} analyzed the spreading dynamics on the temporal contact network, of which the interevent time distribution of the edges exhibits a heavy-tail, {\it i.e.}, it is a temporally inhomogeneous
bursty contact process. They observed that the temporal heterogeneity suppresses the susceptible-infected (SI) spreading dynamics. Later, Cui {\it et al.} \cite{cui2014} found that at the individual level, the burstiness promotes the SI spreading dynamics at the early times and slows down the spreading at the late stages. Recently, Xue {\it at al.} \cite{xue2020} observed that in the temporal network with heterogeneous population, the network temporality can promote or suppress the spreading dynamics, depending on the heterogeneities of population and degree distribution.

Most of the research work of burstiness on temporal network mainly focuses on the contact network without phase transitions. For systems like active matter \cite{bech2016}, whether there is a deep relation between the temporal heterogeneities and the phase transition from disordered state to the ordered (flocking) phase has not been studied yet. Besides, active matter systems, like the Vicsek model \cite{vicsek1995} and the active Brownian particles \cite{roman2012}, are extensively used to describe a wide range of biological
processes, ranging from bacteria movement to animal behavior. A lot of efforts have been spent on novel properties, for instance, the motilit-induced phase separation \cite{cates2012}. However, only a few studies have been carried out on the information or disease spreading dynamics on those systems\cite{nora2020,paolu,forg,zhao}, and little attention has been given on the heterogeneities of the temporal networks that formed by those active matter models.

\begin{figure*}[t]
\includegraphics[width=0.92\linewidth]{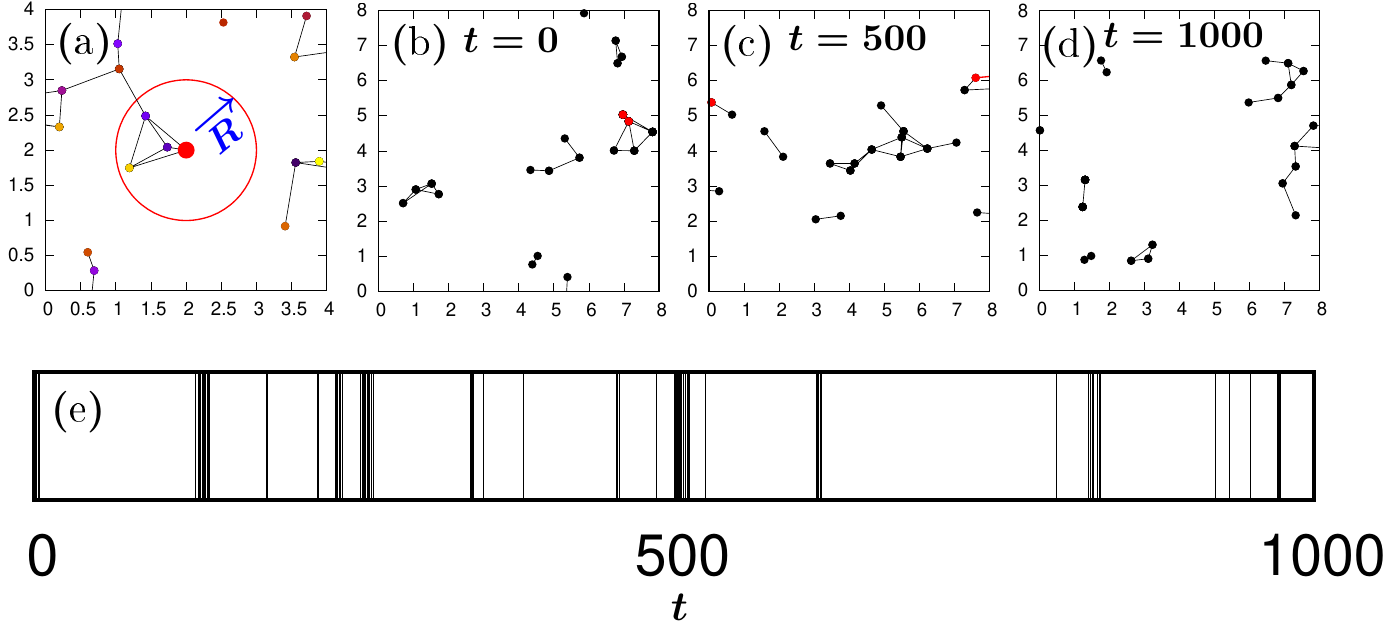} 
\caption{Schematic diagram of the temporal network constructed by the Vicsek model and an example signal. (a) The temporal network construction method, where $R=1$ is the interaction radius. Any pair of particles with their distance less than $R$ is connected with an edge. The snapshots shown in (b)-(d) are instantaneous configurations at $t=0$, $t=500$, and $t=1000$ in a simulation with the system size $L=8$ and particle number $N=20$. The target edge is noted in red, with the full time signal of the target edge for $t<=1000$ is presented in (e). \label{fig1}}
\end{figure*}  
 
In this paper, the burstiness of a specific pair of particles on the temporal network of the Vcisek model is studied numerically. By measuring the distribution of the interevent time for different noise strengths and densities, we find that strong burstiness exists in the whole parameter space. The bursty nature is further characteristic by the burstiness parameters and the memory coefficients. The results show that there is a deep relation between the phase transition of the Vicsek model and the bursty nature of the signals. Besides, the memory plays an unimportant role in the burstiness. Further, the information spreading dynamics on the constructed temporal networks with the susceptible-infected model are analyzed. The results indicate an internal connection between the burstiness and the information spreading speed. 

The organization of this paper is as follows. In Sec. II we introduce the temporal network that are constructed with the Vicsek model. Then, we analyze the burstiness of the interevent times of the temporal network in Sec. III . In Sec. IV we
further discuss the influence of the burstiness to the dynamcis of the information spreading. A short conclusion is given in Sec. V.

\section{Model descriptions \label{sec2} }

In this section, the model system, i.e., the Vicsek model, and the process of temporal network construction are introduced.

\subsection{Vicsek model}

The Vicsek model is one of the best known models that exhibit a phase transition from disordered state to flocks  \cite{ginelli2016}. As proposed by Vicsek and his coauthors \cite{vicsek1995}, with an alignment interaction, the self-propelled particles follow the overdamped dynamics 
\begin{equation}
{\vec x}_i(t+1)={\vec x}_i(t)+{\vec v}_i(t) \Delta t
\end{equation}
where the time unit $\Delta t$ is set to be $1$, ${\vec x}_i(t)$ is the position of the {\it i}-th particle at time $t$, and ${\vec v}_i(t)$ is its velocity. Normally, the velocity is composed of an absolute value $v$, which is set to be $0.5$, and an angle $\theta(t)$, which describes the direction of the velocity. The angle $\theta(t)$ is expressed as
\begin{equation}
\theta(t+1)=\langle \theta(t)\rangle_R+\xi_i(t)
\end{equation}
here, $\langle \theta(t)\rangle_R$ represents the average direction within the interaction radius $R=1$, and the noise $\xi_i(t)$ is a random variable chosen uniformly from the interval $[-\eta \pi,\eta \pi]$, with $\eta$ is the noise strength. The particle density is defined as $\rho=N/L^2$, where $N$ is the particle number and $L$ is the system size. Note that periodic boundary conditions are used for all the simulations in this paper.

\subsection{Temporal network}


Since the Vicsek model was employed to model the dynamics of the active particles, for a target particle $i$, it can only communicate with those particles within the circle of radius $R$. By adopting this simple rule, the temporal network can be naturally constructed as shown in Fig. \ref{fig1} (a). For each pair of particles, an edge is created between them iff their distance is less than $R$.  

In order to study the bursty nature of the temporal network, a specific edge is selected. For an example simulation with system size $L=8$, and particle number $N=20$,
the configurations at $t=0$, $500$, $1000$ are depicted in Fig.~\ref{fig1}(b)-(d). For the target edge (the red edge in Fig.~\ref{fig1}(b)-(c)), the activate signal is recorded (Fig.~\ref{fig1}(e)), and one obtains a series of temporal periods $(t_{b,i},t_{e,i})$, with $t_{b,i}$ and $t_{e,i}$ specifying the beginning and the ending time when the target edge is activated and killed for the $i$th time, respectively. The interevent time is defined as $\tau_i=t_{b,i+1}-t_{e,i}$. Note that for all the data shown below, the system size is $L=32$ and for each data, we have run 200-1000 independent simulations to obtain the averaged results.

\section{Burtiness in the active particle systems  \label{sec3}}

Unlike the well known Poisson process, whose activity pattern is random and its interevent time, $\tau$, follows an exponential distribution, $P_P(\tau)\sim \exp(-\tau/\tau_0)$, the burstiness is described by an interevent time distribution that is different from the purely exponential.

\begin{figure}[t]
\includegraphics[width=0.44\linewidth]{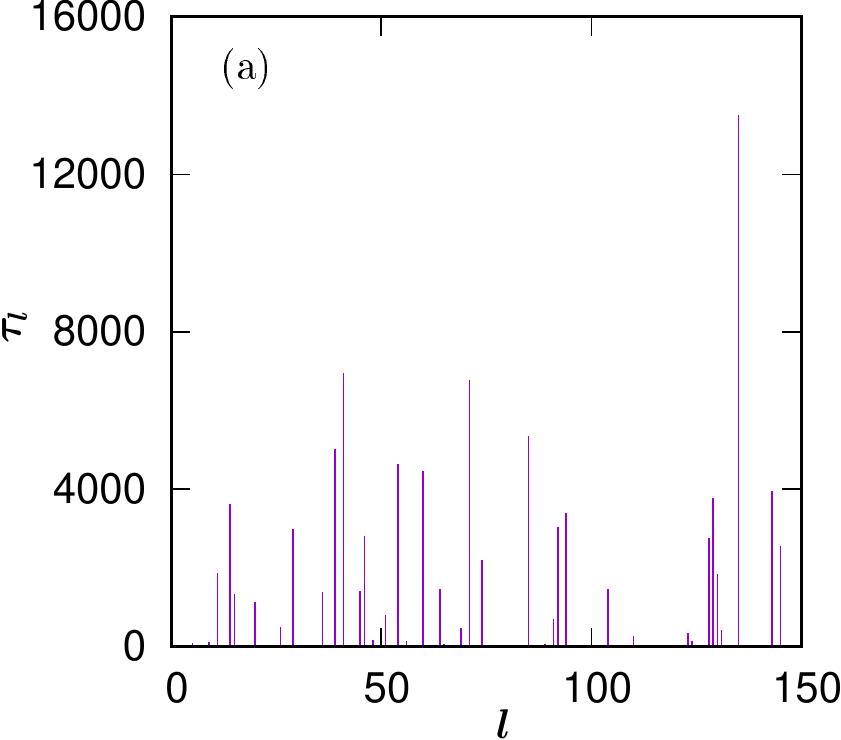} \hspace{3mm}
\includegraphics[width=0.48\linewidth]{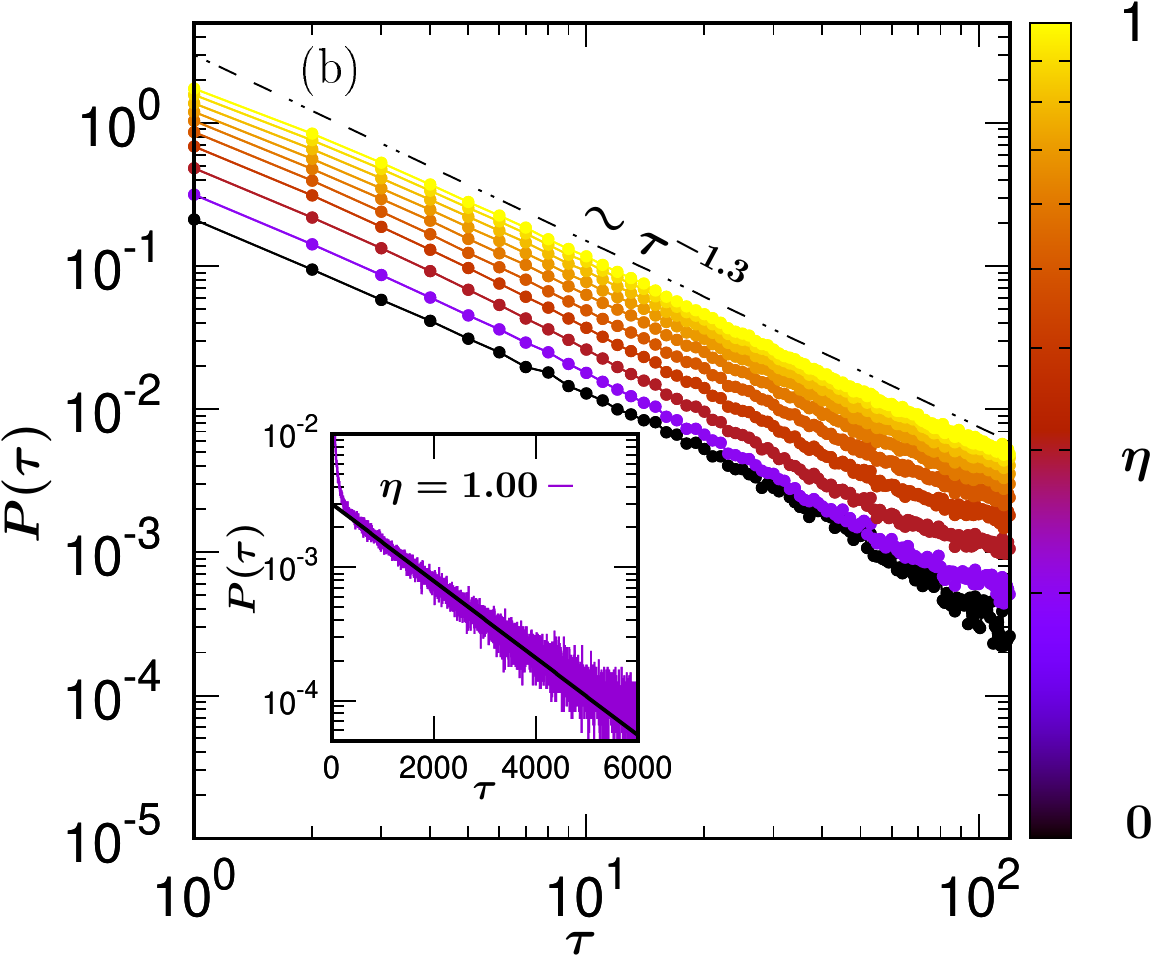} \\
\includegraphics[width=0.41 \linewidth]{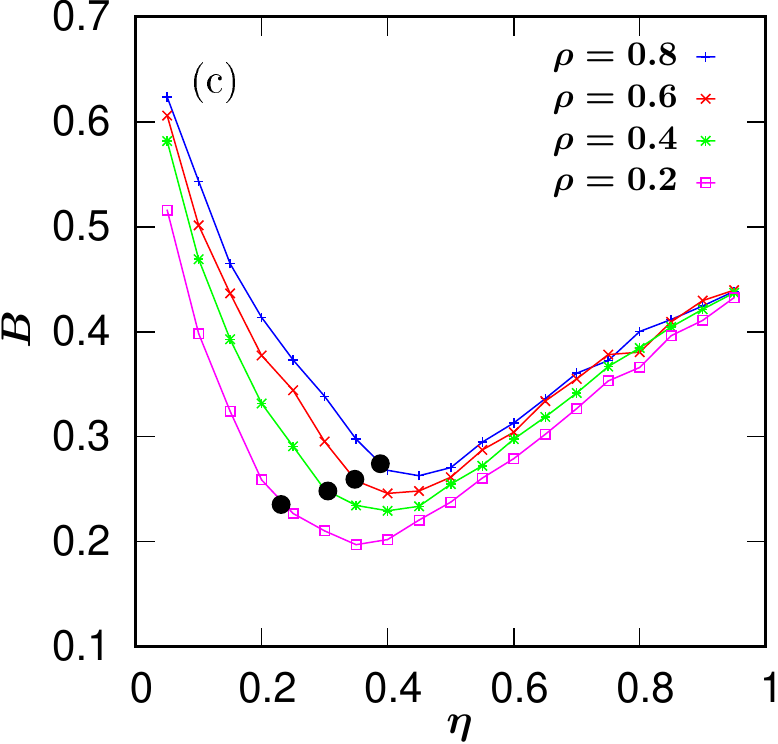} \hspace{5mm}
\includegraphics[width=0.5\linewidth]{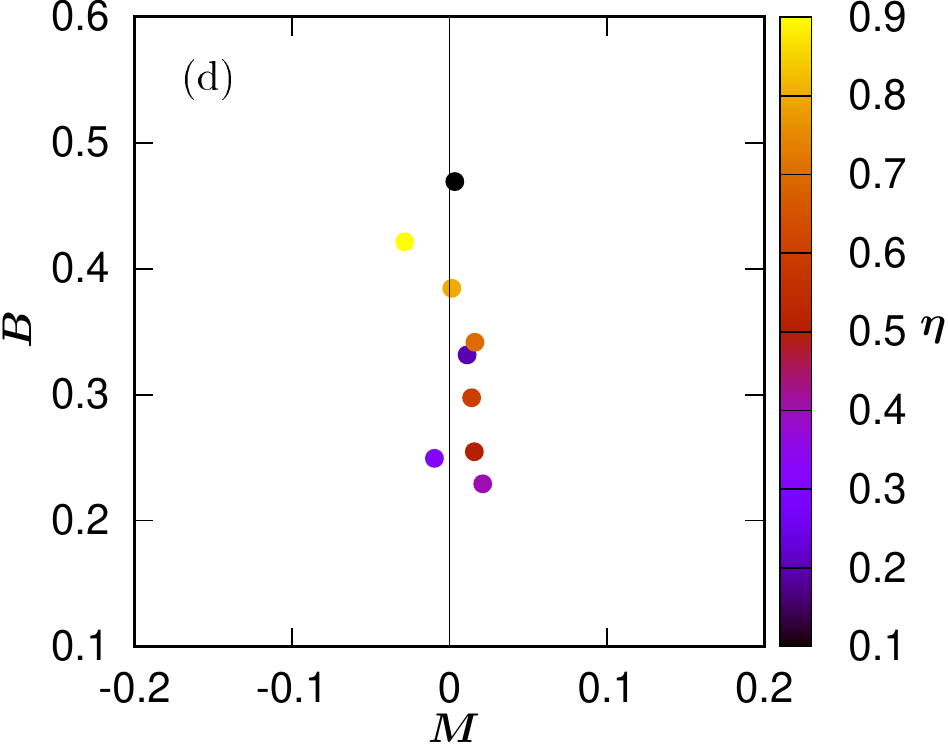}
\caption{ The burstiness nature of the temporal networks. (a) The interevent times recorded in a simulation. (b) The interevent time distribution $P(\tau)$ with $\rho=0.20$ for different noise strength $\eta$. It is shown that for all $\eta$, the interevent time distributions  undergo a power-law decay, and then an exponential decay for very large times (as displayed in the inset). This result can be expressed as $P(\tau) \sim \tau^{-\alpha} \exp(-\frac{\tau}{\tau_E})$, with $\alpha$ the power-law exponent, and $\tau_E$ the characteristic decay time. It indicates that the temporal network formed by the Vicsek model are bursty. (c) The burstiness parameter $B$ for different density $\rho$ and $\eta$. All $B$ are positive confirm that the signals are bursty. Besides, to compare with the location of the critical points (black points, \cite{chate}) reveals that there is a deep connection between the burstiness and the phase transition of the active paricles. (d) The ($M,B$) diagram. The results, which are the same for the human activities \cite{eck,vaz,har}, show that the burstiness parameter $B$ is large and positive, while the memory coefficient $M$ is small and almost negligible. \label{fig2}}
\end{figure}

For the temporal networks constructed by the active particles via the Vicsek model, the signals generated by this temporal network are analyzed. A target edge, which is randomly selected, is tracked after the system reaches its steady state, and the interevent times are recorded. As an example, Fig.~\ref{fig2}(a) shows the first $150$ interevent times for a signal. Using these data, we can easily produce the probability distribution of the interevent times. As proposed by Vazquz {\it et al.} \cite{vazquz2007} that for a burstiness signal, the probability distribution $P(\tau)$ can behave as
 \begin{equation}
 P(\tau)\sim \tau^{-\alpha} \exp\left(-\frac{\tau}{\tau_E} \right)
 \label{eq1}
 \end{equation}
where $\tau_E$ is the characteristic decay time, and $\alpha$ is the exponent. 

To confirm the bursty nature of the temporal network, we carry out simulations for different noise strength and for $\rho=0.20$. In Fig.~\ref{fig2}(b), we find that $P(\tau)$ indeed behave as Eq.~(\ref{eq1}), {\it i.e.}, a power-law decay followed by an exponential cutoff. 

It is unexpected that we have obtained the same $\alpha\approx~1.3$ for different noise strengths. Even though the theoretical understanding of the values of $\alpha$ is lacking, we can still compare our results with other systems. For instance, for the priority queue model with asymptotic power-law distribution of the interevent time, which was first proposed by Barab\'asi \cite{bara}, its power-law exponent $\alpha$ equals to $1$, while for the modifications of this model by considering various detailed factors of human dynamics \cite{gri,gon}, the values of $\alpha=1.5$ \cite{gri} and $\alpha=1.25$ \cite{gon} were observed, respectively. We notice that the value of $\alpha$ for the temporal network of the Vicsek model is close to the values of the modified priority queue models. Nevertheless, more quantities are required to understand the temporal heterogenesis of the temporal network of the Vicsek model.

\subsection{Burstiness parameters}

To further characterize the deviation of the recorded signals from the Poisson process, the burstiness parameter $B$ is calculated as
\begin{equation}
B\equiv \frac{(\sigma_{\tau}/m_\tau-1)}{(\sigma_\tau/m_\tau+1)}=\frac{(\sigma_\tau-m_\tau)}{(\sigma_\tau+m_\tau)}
\end{equation}
where $\sigma_\tau$ and $m_\tau$ are the standard deviation and mean of $P(\tau)$, respectively. 

The burstiness parameter $B\in (-1,1)$. When the signal is strongly bursting, $B=1$; $B=0$ is for the Poisson process; for a completely periodic signal, $B=-1$.
 
The results in Fig.~\ref{fig2}(c) show that positive burstiness parameters exist for different $\eta$ and $\rho$, and interestingly, the burstiness parameters are not monotonically increasing or decreasing. For a specific density, the burstiness parameters decreases for small $\eta$, after they reach a valley value, which is quite close to the critical points of the Vicsek model, the burstiness parameters turn to increase. The nonmonotonic behavior of the burstiness parameter indicates a deep connection between the phase transition of the Vicsek model and the bursty nature of the signals. 

\subsection{Memory coefficients and the $M-B$ diagram}

 Although the burstiness parameter assesses the temporal heterogenesis, it does not characterize possible correlations in the signals. Therefore, another quantity, {\it i.e.}, the memory coefficient, which can further evaluate the memory effect of the bursty signal, is defined as 

\begin{equation}
M\equiv \frac{1}{n_\tau-1} \sum_{i=1}^{n_\tau-1} \frac{(\tau_i-m_1)(\tau_{i+1}-m_2)}{\sigma_1 \sigma_2}
\end{equation}
where $\sigma_1$ ($\sigma_2$) and $m_1$ ($m_2$) are the standard deviation and mean of the first (last) $n_\tau-1$ interevent times. 

With this definition, within a signal, if a short (long) interevent time is more willing to follow another short (long) interevent time, then $M$ is positive. In contrast, a short (long) interevent time followed by a long (short) ones results in a negative $M$. For all $\eta$ and $\rho$ considered, we obtain that $M\approx 0$.

For now, we have empolyed two quantities, {\it i.e.}, the burstiness parameter $B$ and the memory coefficient $M$ to depict the burstiness of the signals. As suggested by Goh and Barab\'asi \cite{goh2008} that to distinguish differect burstiness signals, it is powerful to place the values of $B$ and $M$ for different signals in the ($M,B$) plane, as shown in Fig.~\ref{fig2}(d). To compare with the data shown in Ref. \cite{goh2008}, 
it clearly shows that the results for the temporal network of the active particles are consistent with the contact network from human activities networks \cite{eck,vaz,har}, which have high values of $B$, while a small or negligible value of $M$. 

To summarize, we have found that the burstiness exists in the whole parameter space on the temporal network of the Vicsek model. The deep relation between the burstiness and the phase transition of the Vicsek model is identified from the nonmonotonic behavior of the burstiness parameters. Besides, it is observed that memory plays a negligible role in the burstiness. 

\begin{figure}[t]
\includegraphics[width=0.46\linewidth]{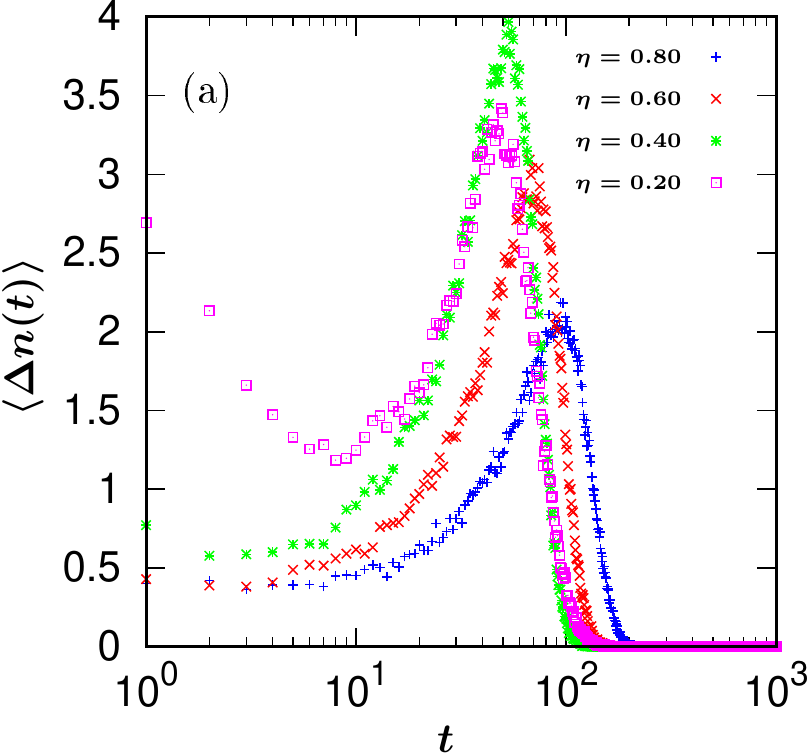} \hspace{3mm}
\includegraphics[width=0.45\linewidth]{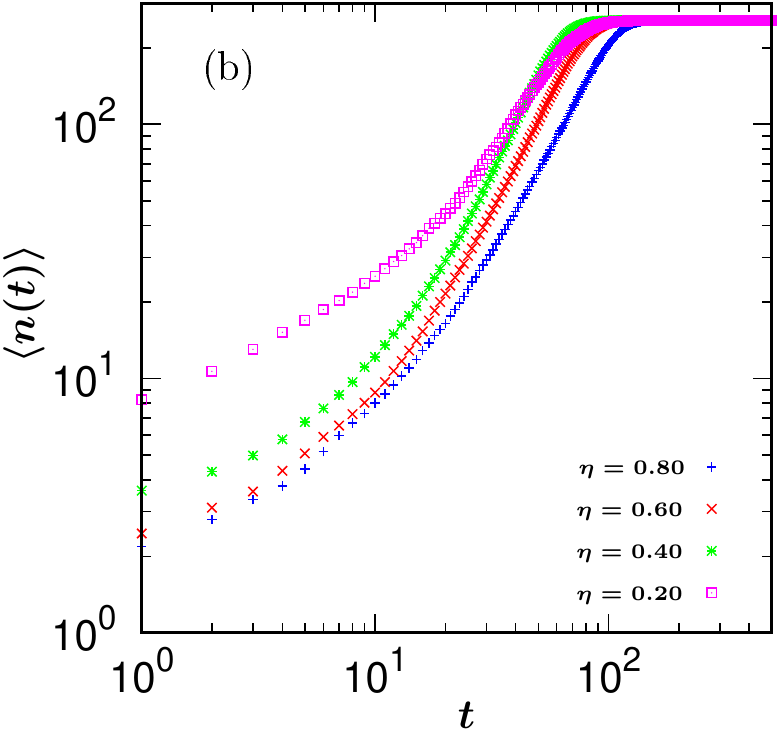} \\
\includegraphics[width=0.45\linewidth]{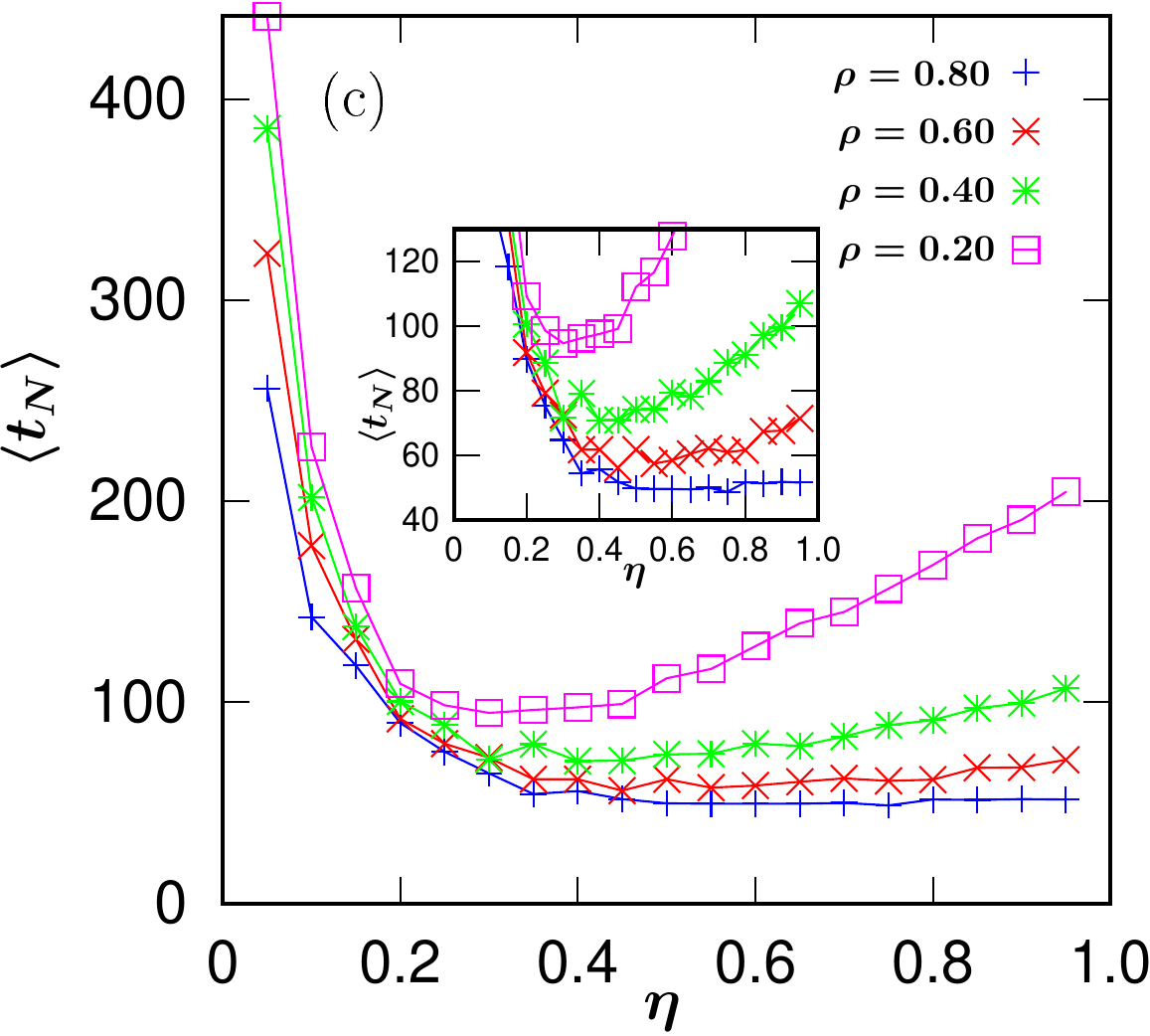} \hspace{5mm}
\includegraphics[width=0.45\linewidth]{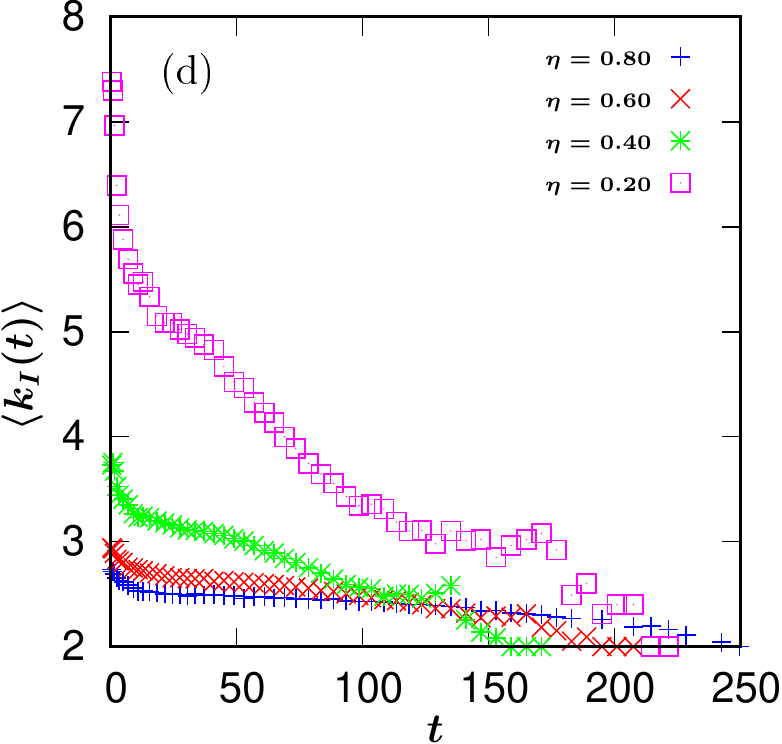}
\caption{The information spreading dynamics in the Vicsek model. (a) The time evolution of the newly infected individuals $\langle \Delta n(t) \rangle$. It is shown that for $\eta=0.20$, at the early stage, $\langle\Delta n(t)\rangle$ decrreases. However, for the rest of the results, this decrease is not exist. (b) The infected density as a function of $t$. (c) The full prevalence time $t_N$ for  different $\rho$ and $\eta$. The location of the valley values are consistence with the burstiness parameters in Fig.~\ref{fig2}(c). (d) The average degree of the newly infected individuals. \label{fig3}}
\end{figure}

\section{Information spreading dynamcis on the temporal network}

As we have mentioned, that the temporal inhomogeneities can strongly affect, {\it i.e.}, promoting or slowing down, the spreading dynamics on the temporal networks \cite{karsai2011,lambiotte2013,horvath2014,
perotti2014,tizzani2018,michalski2020}.  

In order to specify the information spreading dynamics on the temporal networks constructed by the Vicsek model, the simple susceptible-infected (SI) model is employed, where for an active particle, it can only have two states, either susceptible or infected. After the system reaches the steady state, a seed particle is selected to be infected, and for the rest of particles are susceptible. At each time $t$, an infected particle is trying to infect those susceptible particles within the radius of $R$ with the infectious rate $\gamma$, which is set to be $1$ for simplicity. This process will continue until no susceptible individuals left.

In the spreading process, the averaged newly infected density $\langle \Delta n(t)\rangle$ and the infected quantity $\langle n(t) \rangle$ are two key quantities. From Fig.~\ref{fig3}(a), it tells us that at the intermediate and large times ($t\sim (10,500)$) the averaged newly infected density is not monotonically increasing or decreasing, there is a peak value for each $\eta$ demonstrates that the speed of information or disease spreading is accelerated at the early stage, then slows down at later times. At small times ($t \lesssim 10$), for large $\eta$, it increases slowly. However, for small $\eta$ ($\eta=0.20$), $\langle \Delta n(t)\rangle$ decreases from a large value. 

In Fig.~\ref{fig3}(b), we can find that at the early stage ($t\lesssim 20$), the smaller $\eta$ is, the larger number of the individuals $\langle n(t) \rangle$ are infected, which means the faster the information spreads. However, at the intermediate stage, it is complicated. For large times, to evaluate the spreading speed, the full prevalence time $\langle t_N\rangle$, which denotes the full infection on the temporal network, is measured. Fig.~\ref{fig3}(c) shows that for small $\eta$, the full prevalence time is very large indicates the slow spread dynamics. For large $\eta$, when the density increases, $\langle t_N\rangle$ tends to decreases, revealing that the spreading is speeded up by enhancing the density. The similar behavior of $\langle t_N\rangle$ and the burstiness parameter suggests a positive correlation between them. Besides, the full prevalence time has a valley value for each $\rho$, with its location is also consistent with the results shown in Fig.~\ref{fig2}(c).

To further understand the underlying physics of the spreading dynamics, the average degree of the newly infected individuals $\langle k_I(t)\rangle$ are calculated. In Fig.~\ref{fig3}(d), it shows that at the early stage, $\langle k_I(t)\rangle$ for $\eta=0.20$ is much larger than those for the rest of $\eta$. This is identical to the data of $\langle \Delta n(t)\rangle$ at small times. When $\eta$ decreases, $\langle k_I(t)\rangle$ also reduces, illustrating that the smaller $\eta$ is, the faster the information spreads. At large time regime, most of the individuals that belong to large cluster are infected yet, then $\langle k_I(t)\rangle$ tends to zero finally. 

 


\section{Conclusions \label{sec4}}

In this paper, the temporal networks constructed by the Vicsek model are investigated numerically. The interevent time distribution $P(\tau)$ is measured, the results shows that burstiness exists in temporal networks of the active particles, both for small and large noise regime. To further characteristic the burstiness, the burstiness parameter and memory coefficient are calculated. The results show a relation between the burstiness and the phase transition of the Vicsek model. Besides, memory plays a negligible role in the burstiness. Further, the spreading dynamics on the temporal network is analyzed. Our results reveal that a positive correlation between the burstiness and the spreading dynamics is exist. 

Although we have observed some deep connection between the temporal heterogeneities and the phase transition in the active particles from the nonmonotonic bahavior of the burstiness parameter (Fig.~\ref{fig2}(c)) and the full prevalence time (Fig.~\ref{fig3}(c)), the underlying physics is still not clear. It deserves more studies in future.

\section*{Acknowlegement \label{sec5}}

W.Z. acknowledges support from the National Natural Science Foundation of China Youth Fund (Grant No. 12105133) and the Fujian
Provincial Natural Science Foundation of China (Grant No. 2021J011030). Y.D. acknowledges support by the National Natural Science Foundation of China (under Grant No. 12275263), the Science and Technology Committee of Shanghai (under Grant No. 20DZ2210100), and the
National Key R\&D Program of China (under Grant
No. 2018YFA0306501). D.X. is supported
by NNSF of China (Grant No. 12275116) and NSF of Fujian Province of China (Grant No. 2021J02051).


\begin{thebibliography}{11}
\bibitem{holme2012} P. Holme, J. Saramäki, Phys. Rep., {\bf 519}, 97 (2012).
 
\bibitem{holme2015} P. Holme, Eur. Phys. J. B, {\bf 88}, 1 (2015).
 
\bibitem{li2017} A. Li, {\it et al.}, Science, {\bf 358}, 1042 (2017).
 

\bibitem{bara} A. L. Barab\'asi Nature, {\bf 207}, 435 (2005).
 
\bibitem{gold} I. Golding, J. Paulsson, S. M. Zawilski and E. C. Cox, Cell, {\bf 123}, 1025 (2005).

\bibitem{moin} A. Moinet, M. Starnini, and R. Pastor-Satorras, Phys. Rev. Lett., {\bf 114}, 108701 (2015).
 
 
\bibitem{goh2008} K. -I. Goh and A. -L. Barab\'asi, EPL, {\bf 81}, 48002 (2008). 
 
 \bibitem{kim2016} E. -K. Kim and H. -H. Jo, Phys. Rev. E {\bf 94}, 032311 (2016).

\bibitem{moinet2015} A. Moinet, M. Starnini, R. Pastor-Satorras , Phys. Rev. Lett. {\bf 114}, 108701 (2015) 



\bibitem{karsai2011} M.Karsai, {\it et al.}, Phys. Rev. E, {\bf 83}, 025102 (2011).

\bibitem{lambiotte2013} R. Lambiotte, L. Tabourier, and J. C. Delvenne, Eur. Phys. J. B, {\bf 86}, 1 (2013).

\bibitem{horvath2014} D. X. Horv\'ath, and J. Kert\'esz, New J. Phys., {\bf 16}, 073037 (2014).

\bibitem{perotti2014} J. I. Perotti, {\it et al.}, arXiv preprint arXiv:1411.5553.

\bibitem{tizzani2018} M. Tizzani, {\it et al.}, Phys. Rev. E {\bf 98}, 062315 (2018).

\bibitem{michalski2020} R. Michalski, J. Jankowski, and P. Br\'odka, IEEE Access, {\bf 8}, 151208 (2020).



\bibitem{vazquz2007} A. Vazquz {\it et al.}, Phys. Rev. Lett. {\bf 98}, 158702 (2007).

\bibitem{cui2014} A. -X. Cui {it et al.}, Chaos {\bf 24}, 033113 (2014).

\bibitem{xue2020} X. -Y. Xue, {\it et al.}, Chaos {\bf 30}, 113136 (2020).


\bibitem{shae2020} M. R. Shaebani, {\it et al.} Nat. Rev. Phys.
{\bf 2}, 1 (2020).

\bibitem{bech2016} C. Bechinger, {\it et al.}, Rev. Mod. Phys., {\bf 88}, 045006 (2016). 
 
\bibitem{vicsek1995} T. Vicsek, {\it et al.}, Phys. Rev. Lett., {\bf 75}, 1226 (1995).

\bibitem{chate} H. Chat\'e, {\it et al.}, Eur. Phys. J. B {\bf 64}, 451 (2008).

\bibitem{roman2012} P. Romanczuk, {\it et al.}, Eur. Phys. J. Spec. Top., {\bf 202}, 1 (2012).

\bibitem{cates2012} M. E. Cates, and J. Tailleur, Annu. Rev. Condens. Matter Phys., {\bf 6}, 219 (2015).


\bibitem{nora2020} A. Norambuena, F. J. Valencia, and F. Guzmán-Lastra, Scientific Reports, {\bf 10}, 1 (2020).

\bibitem{paolu} M. Paoluzzi, M. Leoni, and M. C. Marchetti, Soft Matter {\bf 16}, 6317 (2020).

\bibitem{forg} P. Forg\'acs, A. Lib\'al, C. Reichhardt, N. Hengartner, and C. J. O. Reichhardt, Sci. Rep. {\bf 12}, 11229 (2022).

\bibitem{zhao} Y. Zhao, C. Huepe, and P. Romanczuk, Sci. Rep. 12, 2588 (2022).

\bibitem{ginelli2016} F. Ginelli, Eur. Phys. J. Spec. Top., {\bf 225}, 2099 (2016). 


\bibitem{gri} G. Grinstein, R. Linsker, Phys. Rev. Lett. {\bf 97}, 130201 (2006). 
\bibitem{gon} B. Gon\c{c}alves, J. J. Ramasco, Phys. Rev. E 78, 026123 (2008).


\bibitem{eck} J. P. Eckmann, E. Moses, and D. Sergi, Proc. Natl.
Acad. Sci. U.S.A., {\bf 101} 14333 (2004). 

\bibitem{vaz} A. V\'{a}zquez, {\it et al.}, Phys. Rev. E, {\bf 73} 036127 (2006). 

\bibitem{har} U. Harder, and M. Paczuski, Physica A, {\bf 361} 329 (2006). 
  





\end{thebibliography}
\end{document}